# Simultaneous Optimization of Signal Timing and Capacity Improvement in Urban Transportation Networks Using Simulated Annealing

*Bahman Moghimi, Navid Kalantari, Camille Kamga, Kyriacos Mouskos*

**Abstract:**

Capacity expansions as well as its reduction have been widely anticipated as important countermeasures for traffic congestion. Although capacity expansion had been traditionally well noticed as a congestion mitigation measure, but it was not until recently that capacity reduction measures such as; congestion pricing, road diet and other such capacity reduction measures were noticed as congestion mitigation measures. Measures such as signal optimization, metering and congestion pricing are intended to affect the travel pattern and assignment behavior of travelers to make the results of the User Equilibrium (UE) traffic assignment, followed by the travelers, closer to the System Optimal (SO) outcomes, intended by the planners. As such, a bi-level optimization model was formulated for the simultaneous optimization of capacity improvement/expansion and signal timing in an urban transportation network. The model takes into account both the effect of higher demand, induced by the capacity expansion, and the rerouting potential effect of traffic signals.

The solution algorithm developed here consists of two major components; the gradient projection algorithm (GP) to solve the lower level traffic assignment problem and the Simulated Annealing (SA) algorithm to solve the master problem. In order to illustrate the method a case study was examined. The analysis and application of the proposed algorithm shows that the performance function gradually converged along the simulation run, and no divergence problem is observed. By applying the developed algorithm, the total network travel time reduced by 13.42%, in which 5.76% is reached by only optimizing the signal times. Using the GP algorithm and taking advantages of dynamic memory in the process of simulation, the whole simulation acquired from computer time is 13.63 seconds with a convergence rate of 0.001.

## 1- Introduction

The complexity of urban transportation problem has led the general practice to decompose it into several, still challenging, sub-problems; some of which are Street Network Design Problem (SNDP) and Signal Setting problem (SSP). This decomposition has made the transportation system planning computationally traceable, and hence it is a compromise between optimality and traceability. The Street Network Design Problem (SNDP), addressed in this study involves the selection of the "near optimal" selection of the best signal timing and roadway capacity expansion simultaneously. Substantial research in the general area of the NDP has been undertaken since the 1970s that dealt with the continuous NDP - where the decision variables are the roadway link capacity expansions, which were considered as continuous. Studies such as those by Abdulaal and LeBlanc (*1979*), Tan et al. (*1979*), Suwansirikul et al. (*1987*), Marcotte (*1983*), Marcotte and Marquis (*1992*), Friesz et al. (*1990*), and Yang and Yagar (*1994*, *1995*) were developed to solve the Continuous Network Design Problem (CNDP). Meanwhile, discrete network design which deals with the decision on whether or not to add/build a certain capacity-expansion-projects have also been considered in the literature. Boyce et al. (*1973*), LeBlanc (*1975*), Jeon et al. (*2006*), Poorzahedy and Abulghasemi (*2005*), Ukkusuri et al. (*2007*), Xiong and Schneider (*1992*), Mouskos (*1991*), Zeng (*1998*), and Gao et. al. (*2005*) are among the most important studies regarding Discrete Network Design Problem (DNDP). The mixed network design problem combines the two CNDP and DNDP and was taken into account in the studies done by Wang and Lo (*2010*), Maganati and Wong (*1984*), Yang and Bell (*1998*), and Luathep et al. (2011).

How to address the signal setting in transportation network has always been a concern to many researchers. The study of Allsop and Charlesworth (*1977*) is among the first studies that approached the signal setting problem; they applied an iterative assignment optimization (IAO) technique. Bi-level programming is another approach which has been considered by many scholars including Suh and Kim (*1992*), Clegg et al. (2001), Meng et al. (*2001*), Wu Sun (*1999*), and Sun and Benekohal (*2005*). Chiou (*2005*; *2007*) used mathematical programming models; Ceylan and Bell (*2005*) applied Genetic Algorithm (GA) to solve the signal setting in the network. Recently, agent-based traffic signal control (List and Mashayekhi, 2016) and self-organizing traffic signal control (Moghimidarzi et al. 2016; Moghimi et al., 2017) are the state-of-the-art simulation-based techniques targeting signal control.

There are few attempts have studied the mixed of signal setting and network design problem. Karoonsoontawong and Waller (*2010*), and Chiou (*2007*), are the ones that combined SSP and CNDP and then formulated them in the format of bi-level programming. In this study, the bi-level approach is also considered. In the lower level the Gradient Projection (GP) is used to make the problem computationally traceable by saving user's paths chosen along the user equilibrium; while in the upper level a meta-heuristic algorithm namely Simulated Annealing is applied to solve the master problem efficiently - due to the non-convexity of the problem. Moreover, the simultaneous optimization of signal timing and capacity expansion can benefit the urban transportation concerns with much lower cost.

The specific problem addressed in this study is:

**Problem Definition**

This research tries to find a near optimal traffic flow, minimizing the total User Equilibrium (UE), in urban transportation network with a fixed demand by a) changing some roadway link expansion, and b) changing signal timing. To do so, the problem is simplified by using a UE static traffic assignment (UESTA) that utilizes the Gradient Projection method (Jayakrishnan, 1994) to assign the vehicles to paths. Meanwhile, only two-phase signal timing is considered to demonstrate the potential of the methodology rather than solve a more realistic problem with more phases. Hence, this study aims to formulate a simplified Urban Network Design Problem (UNDP) by simultaneous optimization of roadway capacity and signal timing as a bi-level problem where i) the upper level is the optimization of the roadway link capacity and the two-phase signal timing and ii) the lower level is the UESTA that estimates the link traffic volume based on the fixed OD traffic demand. This research has been developed a heuristic methodology to solve the problem in the upper level by using Simulated Annealing and the Gradient projection to solve the lower level problem, UESTA. To reach the objectives, the developed model is tested on an urban transportation network.

It is worth noting that in order to develop an algorithm for such complex problem, a simplified version of the model is presented, which has drawbacks. For example, it does not properly model the traffic flow propagation from one origin to one destination as it does not take into consideration the intersection delays in a proper manner (only microscopic traffic simulators have this capability). A dynamic traffic assignment model (DTA) would be more realistic in capturing more properly the traffic flow propagation for all the OD pairs by taking into consideration the time dimension, signal timing and offsets. In addition, the use of the UESTA is not capable to capture properly the impact of expanding one roadway link but not the downstream link as it does not take into consideration the traffic flow propagation from one link to another. In the next section, the formulation of the above stated UNDP is presented.

## 2- Model Formulation

As stated earlier, a bi-level programming is proposed for the simultaneous optimization of capacity improvement/expansion and signal timing in the network. The interaction between link capacity expansion and signal timing is formulized as follows.

$$(\hat{C}, \hat{\varphi}) = \arg\min Z(f^*(C,\varphi), t(f^*, C, \varphi)) \quad [1]$$

$$C, \varphi \in \Lambda$$

$$f \in \Omega$$

Where:

$\hat{C}$ = the vector of optimal capacities of links

$\hat{\varphi}$ = the vector of optimal signal control parameters (e.g. green time)

$f^*$ = the vector of optimal User Equilibrium (UE) link flows

$t$ = the vector of UE link travel times

$\Lambda$ = the set of feasible link capacity and control parameters

$\Omega$ = the set of feasible and equilibrium flows (the User Equilibrium constraint)

The above formulation could be used for the Urban Network Design Problem (UNDP) since it covers two of the four major concerns usually encountered in urban transportation networks which are:

A) The problem of limited capacity enhancement due to traffic engineering efforts such as efficiency improvement (e.g. on-street parking, extra capacity near the intersection, etc.)
B) Signal timing problem at signalized intersection(s)
C) Optimizing the direction and configuration of urban network (e.g. one-way and two-way streets)
D) Congestion pricing

For sake of simplicity, this paper deals with the first two components of the Urban Network Design Problem (UNDP), and also can be reformulated as a bi-level NDP. Each level of the mentioned bi-level programming deals with the role of a specific group of the stakeholders; the upper level represents the objective of the planner(s), while the lower one represents the objectives of the users (e.g. travelers). The upper level of the NDP is the design/supply side transportation which seeks to minimize the network's performance measures (e.g. travel time or travel cost in general). The lower level represents the demand side, which redistributes itself based on the available supply (transportation network characteristics) and is formulated as the user equilibrium (UE) traffic assignment problem.

The transportation network is a graph $G(N, A)$ where $N$ is the set of nodes (e.g. intersections, interchanges) and $A$ is the set of roadway links of the network. We denote the set of signalized intersections as $N_s \subset N$ and the set of candidate links for capacity enhancement as $A_y \subset A$. For each $i \in N_s$ a set of control parameters could be defined as $\varphi_i$. These control parameters may include phasing plan, green time (green split), cycle length and offset. Among the aforementioned controlling parameters, the main one considered in this research is the green time. The traffic control signal is assumed to reduce or increase the capacity proportional to the ratio of green time to the cycle length. This ratio sums up to one for all critical phases of a given signal. Thus, the inclusion of cycle length into this problem is redundant and could be omitted. Formulation (2) is the one used for the bi-level programming of capacity expansion and signal timing in the network.

Upper Level: Minimize the total travel time of the network.

$$\min_{C,\varphi} \sum_{a \in A} x_a^* \cdot t_a(X^*, C, \varphi) \qquad [2]$$

Subject to:

$$\frac{g_{ib}}{\rho_i} + \frac{g_{ib'}}{\rho_i} = 1 \qquad \forall i \in N_s \qquad [2.1]$$

green split-cycle constraint (2.1)

$$(\frac{g}{\rho})_{min} \leq \frac{g_{ib}}{\rho_i} \leq (\frac{g}{\rho})_{max} \qquad \forall i \in N_s, \forall b \in \varphi_i \qquad [2.2]$$

green split lower and upper bounds (2.2)

$$\sum_{a \in A_y} e_a y_a \leq B \qquad \text{Capacity expansion budget constraint} \qquad [2.3]$$

Lower Level (UE Traffic Assignment):

$$Min \sum_{a \in A} \int_0^{x_a} t(w)dw \qquad \text{UE Objective function} \qquad [2.4]$$

Subject to:

$$\sum_{p \in P} f_p^{rs} = q^{rs} \qquad \forall rs \in W \qquad \text{OD demand constraint} \qquad [2.4.1]$$

$$\sum_{p \in P} \sum_{rs} f_p^{rs} \delta_{ap}^{rs} = x_a \qquad \forall a \in A \qquad [2.4.2]$$

Link-path incidence matrix constraints (2.4.2)

$$f_p^{rs} \geq 0 \qquad \forall rs \in W, \quad \forall p \in P \qquad [2.4.3]$$

OD path flow non-negativity constraints (2.4.3)

Where:

$g_{ib}$ = the amount of green time given to the phase 'b' of signal 'i'

$\rho_i$ = the cycle length of signalized intersection $i$

$e_a$ = the unit cost of capacity expansion on link $a$

$y_a$ = the amount of capacity improvement on link $a$

$B$ = the budget level

$f_p^{rs}$ = the flow on route $p$ between origin 'r' and destination 's'

$q^{rs}$ = the demand between $r$ and $s$

$W$ = the set of origin destination pairs

$x_a$ = equilibrium flow in link $a$

$\delta_{ap}^{rs}$ = a 0-1 binary variable which is 1 when link $a$ is on route $p$ between origin 'r' and destination 's' and zero otherwise.

It is difficult to choose one or some projects among a given set of projects so as to optimize certain objective(s) under some resource constraints. For example, for 20 project cases -based on accept/reject decision- there would be $2^{20}$ alternatives networks. Even with excluding half of alternatives due to the resource constraint, still the remaining would be more than 500,000 alternative networks. Therefore, finding an exact solution for a bi-level network design problem is very difficult, because of being NP-hard problem. Other complexity of bi-level network design problem is that it is a combinatorial, non-convex and non-smooth optimization problem.

In this research, the controlling logic of the algorithm searches and tries to come up with an optimum configuration of network -in terms of capacity expansion- and also setting the best signal timing of the network in way that the best match of traffic/demand to supply-changes happens. The best match is the one reaching the lowest traffic delay for all vehicles traveling in the network. Stochastic optimization methods such as meta-heuristic algorithm can be beneficial to give an optimal or a near-optimal solution of such problem. Some of the meta-heuristic algorithms used in transportation network design problem are: Genetic Algorithm (Mathew and Sharma, 2009), Ant System (Poorzahedy and Abulghasemi, 2005), Simulated Annealing (Lee and Yang, 1994), Tabu Search (Drezner and Salhi, 2000), and Particle Swarm Optimization (Zhang and Gao, 2007).

SA is a well-known method for its superior robustness of solving non-convex optimization problems and has been analytically proven to find the optimal solution if used appropriately (*Kirk Patrik,* 1983; *Friesz, 1992; Jahangiri, 2011*). The major shortcoming of SA is that it requires a lot of iterations for finding the optimal solution. Since in this research a simultaneous optimization of capacity expansion and signal timing in the network are concerned, finding a point in the feasible region, solving the UE at each iteration, and then reaching an optimal solution would take much subsequent iterations which could be very time-consuming.

The lower problem – the UE static traffic assignment - the path-based UE traffic assignment Gradient Projection (GP) algorithm (Jayakrishnan, 1994) was used that proved to be extremely fast in this implementation. The Simulated Annealing (SA) and Gradient Projection which are used to solve the NDP are shown in Equation (2).

3- **Solution Algorithm**

The solution algorithm developed here consists of two major components; the GP traffic assignment and SA algorithm. In the following, first, the GP traffic assignment algorithm is presented; then, the implementation of SA on simultaneous optimization of network capacity improvement and signal timing will be discussed.

GP has been originally proposed by Bertsekas (1976). The GP algorithm has been successfully applied to the UE traffic assignment problem by Jayakrishnan (*1994*); then, and further refined by (*Chen and Jayakrishnan, 1998*). The steps of the GP traffic assignment are summarized below as proposed by (*Chen and Jayakrishnan, 1998*)):

Gradient Projection Algorithm for the UE Traffic Assignment (Chen and Jayakrishnan, 1998):

**Step 1**- Path enumeration for each origin-destination pair:

1-1-    Let n=0 and;

$$x_\alpha = 0, \qquad t_\alpha = t_\alpha(x_\alpha) \qquad \forall \alpha \in A \qquad [3]$$

1-2- Find the shortest path, $\bar{k}_{rs}$, between the origin destination pairs r and s.

$$K_{rs} = \{\bar{k}_{rs}\} \qquad [4]$$

1-3- Perform an all-or-nothing assignment

$$f_{\bar{k}_{RS}}^{rs} = q_{rs} \qquad \forall r,s \qquad [5]$$

1-4- Update link flows and travel times based on the all-or-nothing assignment preformed in 1-3.

Step 2- Column generation: find the shortest path between each origin-destination pair based on the current flows on the network and add a new path to the path set $k_{rs}$.

2-1- let *n:= n+1*

2-2- Update link travel times based on current flows:

$$t_\alpha = t_\alpha(x_\alpha) \qquad [6]$$

2-3- Find the shortest path, $\bar{k}_{rs}$, between each origin-destination pair.

2-4- If $\bar{k}_{rs} \in K_{rs}$, let $K_{rs} = K_{rs} \cup \{\bar{k}_{rs}\}$. Otherwise, let the shortest path in $K_{rs}$ be $\bar{k}_{rs}$.

**Step 3**- Traffic assignment:

3-1- Find $d_k^{rs}$ and $S_k^{rs}$, the travel time and travel time derivative respectively for all paths instead of the shortest path.

$$d_k^{rs} = \sum_{\alpha \in A} t_\alpha(x_\alpha) \delta_{\alpha k}^{rs} \qquad \forall k \in K_{rs}$$
$$S_k^{rs} = \sum_{\alpha \in A} t'_\alpha(x_\alpha)(\delta_{\alpha k}^{rs} - \delta_{\alpha \bar{k}_{RS}}^{rs})^2 \qquad \forall k \in K_{rs}, k \neq \bar{k}_{rs} \tag{7}$$

3-2- Update the flows on all non-shortest paths:

$$f_k^{rs} = \max\{[f_k^{rs} - \frac{a}{S_k^{rs}}(d_k^{rs} - d_{\bar{k}_{rs}}^{rs})]\} \qquad \forall rs \in W, \forall k \in K_{rs}, k \neq \bar{k}_{rs} \tag{8}$$

3-3- If $f_k^{rs} = 0$, remove path $k$ from path set $K_{rs}$.

3-4- Update the shortest path flow:

$$f_{\bar{k}_{RS}}^{rs} = q_{rs} - \sum_{\substack{k \in K_{RS} \\ k \neq \bar{k}_{rs}}} f_k^{rs} \tag{9}$$

3-5- Update link flows:

$$x_\alpha = \sum_r \sum_s \sum_{k \in K_{RS}} f_k^{rs} \cdot \delta_{\alpha k}^{rs} \qquad \forall \alpha \in A \tag{10}$$

**Step 4**- Convergence check: if the error is less than the allowable error ε (e.g. 0.001), stop; otherwise, go to Step 2:

$$Err = \max_{rs} \sum_{\substack{k \in K_{rs} \\ k \neq \bar{k}_{rs}}} \frac{f_k^{rs}}{q_{rs}} \left( \frac{d_k^{rs} - d_{\bar{k}_{RS}}^{rs}}{d_k^{rs}} \right) \tag{11}$$

The gradient projection algorithm (GPA) is coupled with SA to solve the proposed NDP. The proposed SA can be summarized as follows:

**Step 1-** Initiation – Base Network Conditions:

1-1- Define initial SA parameters including $T_{int}$, $T_{fin}$, $L$, and α.

1-2- Perform GPA and find the network total travel time – the value of upper level objective function - for the base condition.

1-3- Find the utility of each project based on its corresponding network travel time and sort them in descending order.

1-4- Find the initial transportation network solution (network state: roadway link capacities and signal timing) by selecting the projects with the highest utility such that the Budget constraint is not violated. Also, assign an initial green split ($g_i/\rho_i$) of 0.5 to all signalized intersections.

1-5- Name the initial solution vector *i*.

1-6- Perform GPA and find the value of objective function signified as $E_i$, and set $E_{best} = E_i$.

1-7- Set *n=0*.

1-8- Put the value of $T_{int}$ into $T_n$.

**Step 2**- Neighborhood search:

2-1- Randomly choose a decision variable (a link or an intersection), such that, in each iteration only one change is being made from the current solution set: either a change in link capacity or a change in timing of a signalized intersection.

2-2- If in the previous step (2-1), link is chosen, then randomly pick one link out of set of links; otherwise, do so for signals. Name the picked link or signal *h*.

2-3- If *h* is an intersection, come up with a new traffic control parameter as follows:

$$\frac{g_{hb}}{\rho_h} = 1 - \frac{g_{hb'}}{\rho_h} = Rand\ [(\frac{g}{\rho})_{min}\ ,\ (\frac{g}{\rho})_{max}] \qquad [12]$$

Then go to 2-5.

2-4- If *h* is a link and if the sum of the costs ($\sum_{a \in A_y} e_a$) of all the capacitated links are below the budget level (B), improve the capacity of the selected link and go to 2-5; otherwise, reject the randomly selected capacitated link and go to 2-5.

2-5- Perform GPA and find the value of objective function (network travel time) denoted by $E_j$.

2-6- Measure the amount of change in the objective function as $\Delta E_{ij} = E_j - E_i$

**Step 3**- Update the current and the optimum solution. If $\Delta E_{ij} \geq 0$, put $i=j$; also if $E_j \leq E_{best}$, put $E_j$ into $E_{best}$ and $E_{best} = E_j$; otherwise, find a random number, *ra*, and if $\exp\left(\dfrac{\Delta E_{ij}}{T_n}\right) \geq ra$, then $i=j$ and go to step 4; otherwise, go to step 2.

**Step 4**- If $n < L$, then let $n := n+1$ and go to step 2; otherwise, $n=0$ and go to step 5.

**Step 5**- Convergence check:

5-1- Cooling the temperature by $T_{n+1} < \alpha . T_n$.

5-2- If $T_{n+1} < T_{fin}$, go to the next step; otherwise, go to step 2.

5-3- Report $E_{best}$ and $F_{best}$ as the best objective value and the solution vector, respectively.

The model and solution algorithm presented here is applied on a test network which is discussed in the next section.

## 4- Model Implementation and Result Discussion

The network used in this work as a case study consists of 32 links, 5 signalized intersections and 12 Origin-Destination (O/D) pairs. This network is shown in Figure (1), while the O/D matrix is depicted in Table (1). The shadowed nodes in Figure 1 are the signalized intersections. The volume-delay function employed here is the well-known Bureau of Public Road (BPR) cost function. In addition to the BPR cost function, the Webster-delay function, shown in Equation (13), should be added to all links entering signalized intersections.

$$W = 0.9 \left\{ \dfrac{\rho_a \left[1 - \left(\dfrac{g_{ab}}{\rho_a}\right)\right]^2}{2\left[1 - \left(\dfrac{x_a}{S_a}\right)\right]} + \dfrac{\left(\dfrac{x_a}{S_a\left(\dfrac{g_{ab}}{\rho_a}\right)}\right)^2}{2x_a\left[1 - \dfrac{x_a}{S_a\left(\dfrac{g_{ab}}{\rho_a}\right)}\right]} \right\} \quad [13]$$

In equation (13), $S_a$ is the practical capacity of link *a*, and all other variables are the same as defined previously. As the cost function given by Webster is not continuous when it approaches to the practical capacity of the link, the presented signal's volume-delay function developed by Van Vuren and VanVliet (*1992*) is adopted which is shown in Equation (14):

$$\begin{aligned} &\text{for} \quad x \leq \hat{x}_{kirk} = s.\lambda - \sqrt{\frac{s.\lambda}{T}} \quad \text{then} \\ &w = w_1 + w_2 = \frac{CL.\ s(1-\lambda)^2}{2(s-x)} + \frac{x}{2s.\lambda\ (s.\lambda - x)} \\ &\text{for} \quad x \geq \hat{x}_{kirk} \quad \text{then} \\ &w = w_1 + w_2 + w_{od} = (x - \hat{x}_{kirk})\frac{T}{2s.\lambda} + w(\hat{x}_{kirk}) \\ &\lambda = \frac{g_{ab}}{\rho_a} \end{aligned} \quad [14]$$

where $x$ is the link's flow, $s$ is the saturation capacity of intersection, $\lambda$ is the green time ratio, and T is the duration of time for traffic study (the reasonable time interval of 60 min is used in this study).

The information regarding links of the network is given in Table (2) which includes the initial capacity, free travel time, cycle length, and capacity expansion. The term "link division" in Table 2 indicates whether that link enters a signalized intersection or not. If the value is 1, the link enters a signalized intersection and subsequently the signal's volume-delay function will be added to its link's BPR function. It should be noted that all links are considered as candidates for capacity expansion. The allowable measure for link capacity expansion is 30% of its initial value. In another word, if a link is selected to be expanded, its capacity will increase by 130% of its initial capacity; otherwise, the capacity will remain the same. The corresponding unit-cost for such capacity expansion of each link is given in Table 2. The total cost required for improving all the links is 2208 units. Thereby, the budget constraint is set to be 900 units, representing about 40% of the total cost.

Regarding SA parameters, the cooling rate was set at 0.99 and at each temperature level, 30 iterations of were simulated. 11700 iterations of SA were performed in order to find the optimal solution of the test network. Table (3) shows the results of applying the SA algorithm to the network. In Figure (2), the highlighted links in thicker colors are the ones that should be capacitated to benefit a better performance of such network given the budget level.

*Main Results of the test network:*

The original network travel time was 7615000 seconds and the corresponding SA-UNDP implementation resulted in a total network travel time became 6440450 seconds, a reduction of 13.42%, out of which 5.76% is due to optimizing the signal timing. Figure (3) depicts the SA-UNDP convergence performance.

The computational time to execute the SA-UNDP for this test network is 13.63 seconds to reach a 0.001 convergence criterion value. Figure (4) displays the sensitivity analysis of the problem with respect to the budget level. As expected, the change of the objective function has improved as the budget level is increased. However, adding a budget over 50% did not considerably improve the whole travel time in the network.

## 5- Summary and Conclusions

The problem addressed in this study is the simultaneous optimization of the urban Network Design Problem (UNDP) signal timing and link capacity expansion using the Simulated Annealing (SA) metaheuristic procedure. The specific UNDP addressed in this study is a simplification of the actual UNDP where the signal timing is restricted to two phases and each roadway link is allowed to be physically expanded by 30% or 0% - Hence the near-optimal signal timing is the corresponding green split of each signalized intersection and the near-optimal infrastructure network configuration is the set of roadway links that have been expanded by 30% given that the Budget constraint is not violated. The total network travel time of the network at each iteration was estimated using the static User Equilibrium (UE) traffic assignment Gradient-Projection (GP) model. The developed SA-UNDP procedure has been tested on a small test network to demonstrate the methodology rather than demonstrate its implementation on larger realistic networks.

The main results of the SA-UNDP on the test network are: The number of SA iterations is 11700 at a convergence criterion value of 0.001. The corresponding computational time on a Pentium four PC computer was 13.63 seconds. The reduction in total network travel time from the original base network signal timing and network configuration at convergence is 13.42%, which provides confidence that the procedure could be used successfully for larger networks. The sensitivity analysis on the budget constraint demonstrated that a budget of up to 50% of the total potential budget results in a maximum reduction in total travel time; where the expansion of additional roadway links does not produce additional benefits.

The developed SA-UNDP can be further improved to handle more signal timing phases, including offsets, which limit the effectiveness of this model. In addition, the static traffic assignment can be replaced with a dynamic traffic assignment that will produce more realistic emulation of the traffic flow propagation as it takes explicitly the signal timing including offsets. Such additional improvements while they will make the procedure more capable to handle realistic networks, they will increase the computational performance of the model substantially. Further, other metaheuristics may be explored such that the most efficient procedure could be developed to handle more realistic networks.

## 6- Reference:

Table (1): Origin destination matrix

|    | 1   | 4   | 5   | 9   | 10  | 12 | 13  |
|----|-----|-----|-----|-----|-----|----|-----|
| 1  | –   | 200 | –   | 500 | –   | –  | –   |
| 2  | –   | –   | –   | 300 | –   | –  | –   |
| 4  | –   | –   | –   | –   | 300 | –  | –   |
| 5  | 200 | –   | –   | 200 | 300 | –  | 300 |
| 9  | 200 | –   | –   | –   | –   | –  | –   |
| 10 | –   | –   | 400 | –   | –   | –  | –   |
| 12 | –   | –   | –   | –   | –   | –  | 400 |
| 13 | 300 | –   | –   | –   | –   | –  | –   |

Table (2): The specifications of links

| Origin | Destination | Free travel time (s) | Capacity | Link division condition | Cycle length | Green Ratio | Same-phase Node | Capacity Expansion Condition | Capacity Expansion | Unit Cost | X-kink |
|--------|-------------|----------------------|----------|-------------------------|--------------|-------------|-----------------|------------------------------|--------------------|-----------|--------|
| 1 | 3 | 450 | 800  | 1  | 100 | 0.5 | 7  | 1 | 266.7 | 67 | 399.3 |
| 2 | 3 | 450 | 700  | 1  | 100 | 0.5 | 4  | 1 | 233.3 | 58 | 349.4 |
| 2 | 8 | 530 | 800  | 1  | 110 | 0.5 | 10 | 1 | 266.7 | 67 | 399.3 |
| 3 | 1 | 550 | 800  | -1 | 0   | 0   | 0  | 1 | 266.7 | 67 | 0     |
| 3 | 2 | 550 | 700  | -1 | 0   | 0   | 0  | 1 | 233.3 | 58 | 0     |
| 3 | 4 | 290 | 400  | -1 | 0   | 0   | 0  | 1 | 133.3 | 33 | 0     |
| 3 | 7 | 620 | 1000 | 1  | 120 | 0.5 | 11 | 1 | 333.3 | 83 | 499.3 |
| 4 | 3 | 450 | 400  | 1  | 100 | 0.5 | 2  | 1 | 133.3 | 33 | 199.5 |
| 4 | 6 | 560 | 700 | 1 | 90 | 0.5 | 12 | 1 | 233.3 | 58 | 349.4 |
| 5 | 6 | 550 | 1100 | 1 | 90 | 0.5 | 7 | 1 | 366.7 | 92 | 549.2 |
| 6 | 4 | 320 | 700 | -1 | 0 | 0 | 0 | 1 | 233.3 | 58 | 0 |
| 6 | 5 | 530 | 500 | -1 | 0 | 0 | 0 | 1 | 166.7 | 42 | 0 |
| 6 | 7 | 630 | 900 | 1 | 120 | 0.5 | 8 | 1 | 300 | 75 | 449.3 |
| 6 | 12 | 450 | 600 | -1 | 0 | 0 | 0 | 1 | 200 | 50 | 0 |
| 7 | 3 | 650 | 1000 | 1 | 100 | 0.5 | 1 | 1 | 333.3 | 83 | 499.3 |
| 7 | 6 | 520 | 900 | 1 | 90 | 0.5 | 5 | 1 | 300 | 75 | 449.3 |
| 7 | 8 | 490 | 1000 | 1 | 110 | 0.5 | 9 | 1 | 333.3 | 83 | 499.3 |
| 7 | 11 | 530 | 800 | 1 | 105 | 0.5 | 13 | 1 | 266.7 | 67 | 399.3 |
| 8 | 2 | 430 | 800 | -1 | 0 | 0 | 0 | 1 | 266.7 | 67 | 0 |
| 8 | 7 | 440 | 1000 | 1 | 120 | 0.5 | 6 | 1 | 333.3 | 83 | 499.3 |
| 8 | 9 | 290 | 1200 | -1 | 0 | 0 | 0 | 1 | 400 | 100 | 0 |
| 8 | 10 | 550 | 1000 | -1 | 0 | 0 | 0 | 1 | 333.3 | 83 | 0 |
| 9 | 8 | 450 | 1000 | 1 | 110 | 0.5 | 7 | 1 | 333.3 | 83 | 499.3 |
| 10 | 8 | 450 | 1000 | 1 | 110 | 0.5 | 2 | 1 | 333.3 | 83 | 499.3 |
| 10 | 11 | 430 | 1000 | 1 | 105 | 0.5 | 12 | 1 | 333.3 | 83 | 499.3 |
| 11 | 7 | 550 | 800 | 1 | 120 | 0.5 | 3 | 1 | 266.7 | 67 | 399.3 |
| 11 | 10 | 450 | 1000 | -1 | 0 | 0 | 0 | 1 | 333.3 | 83 | 0 |
| 11 | 12 | 240 | 1000 | -1 | 0 | 0 | 0 | 1 | 333.3 | 83 | 0 |
| 11 | 13 | 690 | 700 | -1 | 0 | 0 | 0 | 1 | 233.3 | 58 | 0 |
| 12 | 6 | 550 | 500 | 1 | 90 | 0.5 | 4 | 1 | 166.7 | 42 | 249.5 |
| 12 | 11 | 250 | 1000 | 1 | 105 | 0.5 | 10 | 1 | 333.3 | 83 | 499.3 |
| 13 | 11 | 480 | 700 | 1 | 105 | 0.5 | 7 | 1 | 233.3 | 58 | 349.4 |

Table (3): The optimum measures of green ratios and expanded capacities after SA implementation.

| Origin | Destination | Expanded Capacity | Cycle Length | Green Ratio | Origin | Destination | Expanded Capacity | Cycle Length | Green Ratio |
|---|---|---|---|---|---|---|---|---|---|
| 1 | 3 | 267 | 100 | 0.61 | 7 | 8 | 0 | 110 | 0.38 |
| 2 | 3 | 0 | 100 | 0.39 | 7 | 11 | 0 | 105 | 0.22 |
| 2 | 8 | 267 | 110 | 0.62 | 8 | 2 | 0 | 0 | 0 |
| 3 | 1 | 267 | 0 | 0 | 8 | 7 | 0 | 120 | 0.45 |
| 3 | 2 | 0 | 0 | 0 | 8 | 9 | 0 | 0 | 0 |
| 3 | 4 | 133 | 0 | 0 | 8 | 10 | 0 | 0 | 0 |
| 3 | 7 | 333 | 120 | 0.55 | 9 | 8 | 0 | 110 | 0.38 |
| 4 | 3 | 133 | 100 | 0.39 | 10 | 8 | 0 | 110 | 0.62 |
| 4 | 6 | 0 | 90 | 0.29 | 10 | 11 | 0 | 105 | 0.78 |
| 5 | 6 | 367 | 90 | 0.71 | 11 | 7 | 0 | 120 | 0.55 |
| 6 | 4 | 233 | 0 | 0 | 11 | 10 | 0 | 0 | 0 |
| 6 | 5 | 167 | 0 | 0 | 11 | 12 | 0 | 0 | 0 |
| 6 | 7 | 300 | 120 | 0.45 | 11 | 13 | 233 | 0 | 0 |
| 6 | 12 | 200 | 0 | 0 | 12 | 6 | 0 | 90 | 0.29 |
| 7 | 3 | 0 | 100 | 0.61 | 12 | 11 | 333 | 105 | 0.78 |
| 7 | 6 | 0 | 90 | 0.71 | 13 | 11 | 233 | 105 | 0.22 |

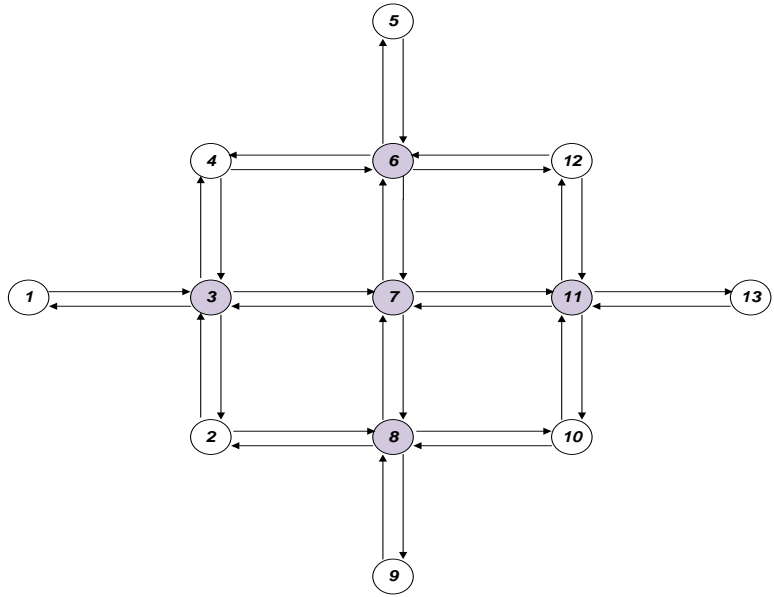

Figure (1): Network with 32 link and 5 signalized intersections

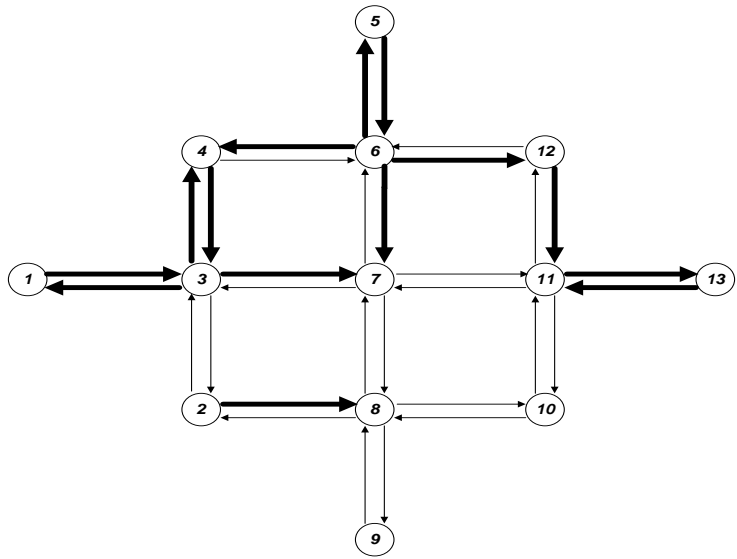

Figure (2): links whose capacities should be increased (links with darker colors).

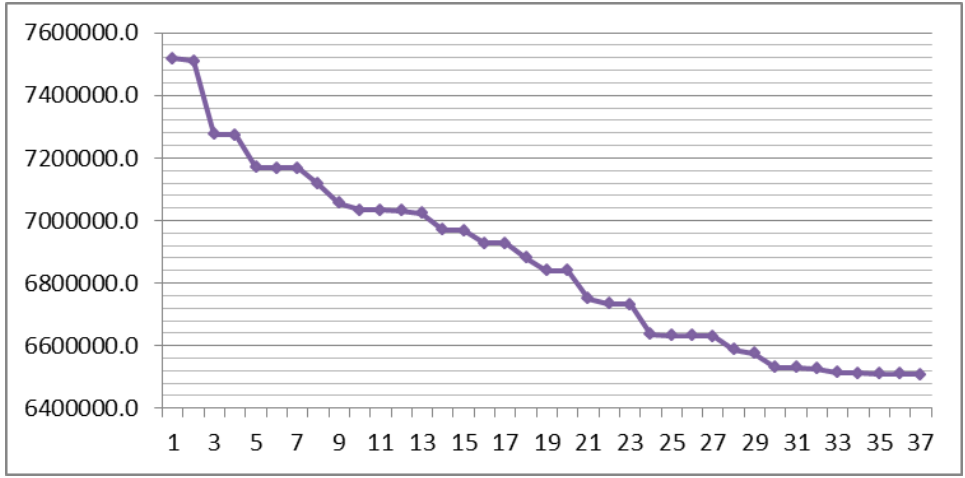

Figure (3): The convergence of the proposed algorithm.

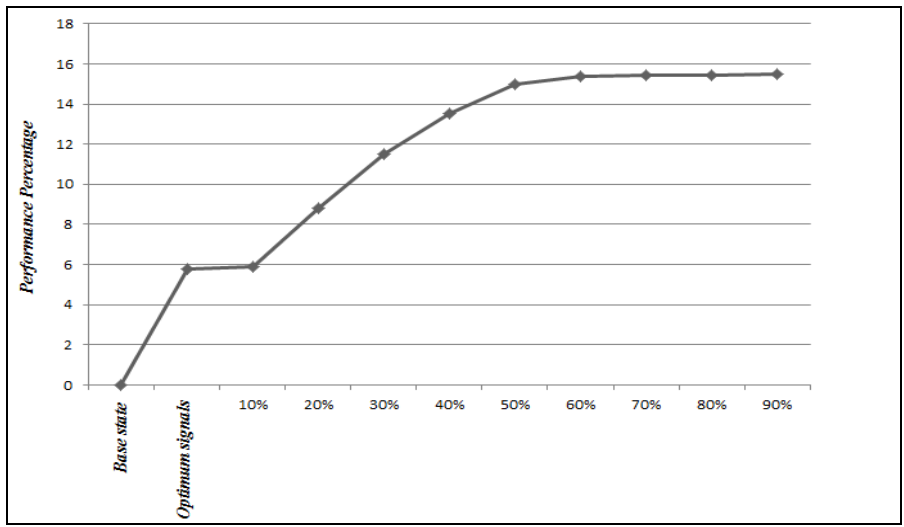

Figure (4): Sensitivity analysis of problem with respect to the budget constraint.